\documentclass[sigconf]{aamas}

\usepackage{balance} 

\makeatletter
\gdef\@copyrightpermission{
  \begin{minipage}{0.2\columnwidth}
   \href{https://creativecommons.org/licenses/by/4.0/}{\includegraphics[width=0.90\textwidth]{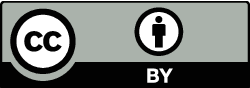}}
  \end{minipage}\hfill
  \begin{minipage}{0.8\columnwidth}
   \href{https://creativecommons.org/licenses/by/4.0/}{This work is licensed under a Creative Commons Attribution International 4.0 License.}
  \end{minipage}
  \vspace{5pt}
}
\makeatother

\setcopyright{ifaamas}
\acmConference[AAMAS '25]{Proc.\@ of the 24th International Conference
on Autonomous Agents and Multiagent Systems (AAMAS 2025)}{May 19 -- 23, 2025}
{Detroit, Michigan, USA}{Y.~Vorobeychik, S.~Das, A.~Nowé  (eds.)}
\copyrightyear{2025}
\acmYear{2025}
\acmDOI{}
\acmPrice{}
\acmISBN{}

\acmSubmissionID{7}

\title{Empirical Hardness in Multi-Agent Pathfinding: Research Challenges and Opportunities}

\subtitle{Blue Sky Ideas Track}

\author{Jingyao Ren}
\affiliation{
  \institution{University of Southern California}
  \city{Los Angeles, CA}
  \country{United States}}
\email{jingyaor@usc.edu}

\author{Eric Ewing}
\affiliation{
  \institution{Brown University}
  \city{Providence, RI}
  \country{United States}}
\email{eric_ewing@brown.edu}

\author{T. K. Satish Kumar}
\affiliation{
  \institution{University of Southern California}
  \city{Los Angeles, CA}
  \country{United States}}
\email{tkskwork@gmail.com}

\author{Sven Koenig}
\affiliation{
  \institution{University of California, Irvine}
  \city{Irvine, CA}
  \country{United States}}
\email{sven.koenig@uci.edu}

\author{Nora Ayanian}
\affiliation{
  \institution{Brown University}
  \city{Providence, RI}
  \country{United States}}
\email{nora_ayanian@brown.edu}

\begin{abstract}
Multi-agent pathfinding~(MAPF) is the problem of finding collision-free paths for a team of agents on a map. Although MAPF is NP-hard, the hardness of solving individual instances varies significantly, revealing a gap between theoretical complexity and actual hardness. This paper outlines three key research challenges in MAPF empirical hardness to understand such phenomena. The first challenge, known as algorithm selection, is determining the best-performing algorithms for a given instance. The second challenge is understanding the key instance features that affect MAPF empirical hardness, such as structural properties like phase transition and backbone/backdoor. The third challenge is how to leverage our knowledge of MAPF empirical hardness to effectively generate hard MAPF instances or diverse benchmark datasets. This work establishes a foundation for future empirical hardness research and encourages deeper investigation into these promising and underexplored areas.

\end{abstract}


\newcommand{\BibTeX}{\rm B\kern-.05em{\sc i\kern-.025em b}\kern-.08em\TeX}

\begin{document}


\pagestyle{fancy}
\fancyhead{}


\maketitle 


\section{Introduction}
\begin{figure}[t]
    \centering
    \includegraphics[width=\columnwidth]{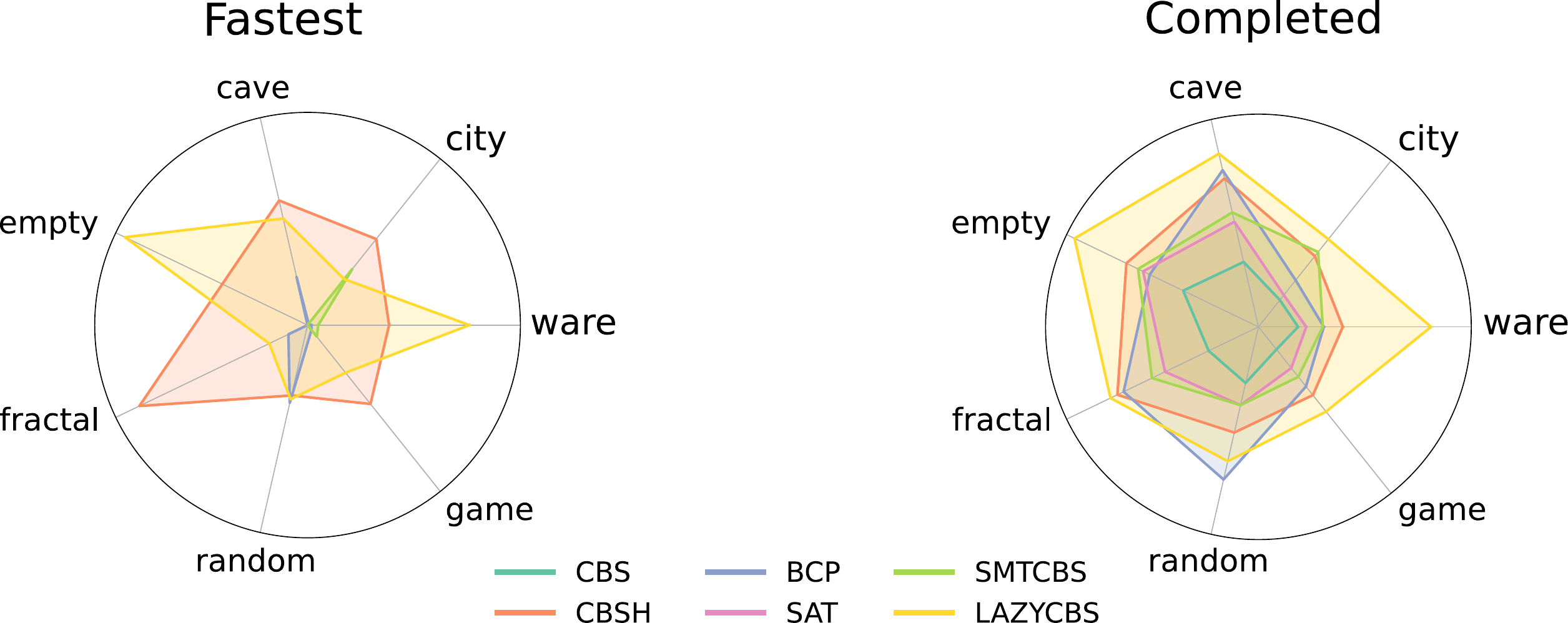}
    \caption{Number of instances where different MAPF algorithms are the fastest in runtime or successfully completed within the time limit across various map types.} 
    \Description{A figure showing the number of instances where different MAPF algorithms are the fastest in runtime or successfully completed within the time limit across various map types.}
    \label{fig:spider}
\end{figure}

Multi-agent pathfinding (MAPF) is the problem of finding collision-free paths for a team of agents from their respective start to goal locations in a shared environment, such as a grid map~\cite{stern2019mapf}. This fundamental problem in artificial intelligence and robotics has numerous real-world applications, such as automated warehouses~\cite{li2021lifelong, ma2019lifelong, honig2019persistent}, trajectory planning for unmanned-aerial-vehicle (UAV)~\cite{honig2018trajectory}, and swarm control~\cite{li2020moving}. Solving MAPF efficiently is crucial for designing scalable and reliable real-world multi-agent systems.

Solving MAPF problems optimally for makespan or sum-of-cost is NP-hard~\cite{yu2013structure}, even for planar graphs~\cite{yu2015intractability} and grid-based problems~\cite{banfi2017intractability}. This means that the worst-case running times grow exponentially with the problem size, such as the total number of agents. However, in practice, real-world MAPF instances are often solved fairly quickly by optimal algorithms~\cite{ewing2022betweenness, shen2023tracking}. This discrepancy highlights a critical gap: while MAPF 
problems are challenging in theory, the hardness of individual instances varies significantly in practice.
Understanding the factors that cause the significant variance of instance hardness is essential but remains poorly understood.

This gap has led to the research of \emph{Empirical Hardness}, which studies the actual difficulty of solving specific problem instances. Empirical hardness is often measured by the runtime an algorithm takes to solve a given instance~\footnote{Other metrics, like memory usage or algorithm-specific metrics (e.g., number of DP calls~\cite{mitchell1992hard} or node expansions~\cite{ren2024map}), are also used to measure empirical hardness.}. Empirical hardness research seeks to understand how problem features correlate with instance hardness and leverage these insights to improve algorithm performance. It has achieved significant success in many challenging problems such as propositional satisfiability (SAT)~\cite{cheeseman1991really}, traveling salesman problem (TSP)~\cite{gent1996tsp}, and combinatorial auctions~\cite{leyton2006empirical}. 

\begin{figure*}[t]
    \centering
        \centering
        \includegraphics[width=.8\linewidth]{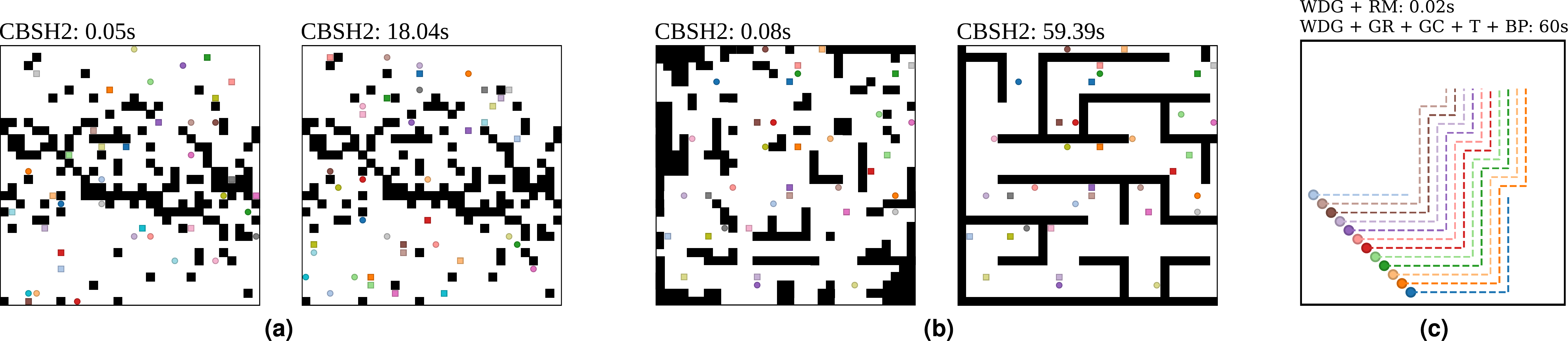}
    \caption{Runtime for CBSH2-RTC~\cite{li2021pairwise} algorithm for different MAPF instances. Circles and squares indicate start/goal locations, and obstacles are in black. (a) MAPF instances on the same map, the same number of agents, but significantly different runtimes. (b) MAPF instances with the same number of obstacles and same start/goal locations but dramatically different runtimes. (c) Runtime variation of CBSH2-RTC algorithm using different combinations of heuristics, dashed lines show planned paths.\protect\footnotemark}
    \Description{A figure showing that different MAPF instances, even with the same number of agents and obstacles, can still have dramatically different empirical hardness in terms of runtime.}
    \label{fig:diff_instances}
\end{figure*}

The empirical hardness of MAPF is a less studied topic. Most existing MAPF research has focused on algorithm design and heuristic enhancements. However, understanding instance-specific hardness variations is critical for developing new algorithms. As shown in Figure~\ref{fig:spider}, MAPF algorithms exhibit varying strengths and weaknesses across different instances, with no single algorithm consistently outperforming others in all instances. How can we select the best algorithm for different instances? Furthermore, even the same algorithm can exhibit significant runtime differences for instances with the same size (e.g., Figure~\ref{fig:diff_instances}(a)(b)). What makes such differences in instance hardness? How can we make use of this knowledge? This paper aims to explore these questions, highlighting key challenges and opportunities in understanding and leveraging the empirical hardness of MAPF.
\footnotetext{We thank Tzvika Geft for providing the corner case illustrated in Figure~\ref{fig:diff_instances}(c).}

\section{Challenge I: When to Choose Which Algorithm?}
As shown in Figure~\ref{fig:spider}, there is no single MAPF algorithm that consistently outperforms others in all cases~\cite{ewing2022betweenness, kaduri2021experimental}. Thus, selecting the fastest algorithm on a case-by-case basis is crucial and can significantly reduce overall runtime~\cite{ren2021mapfast, kaduri2020algorithm}. This area of research is known as algorithm selection and configuration.

\begin{figure*}[t]
    \centering
        \includegraphics[width=.8\linewidth]{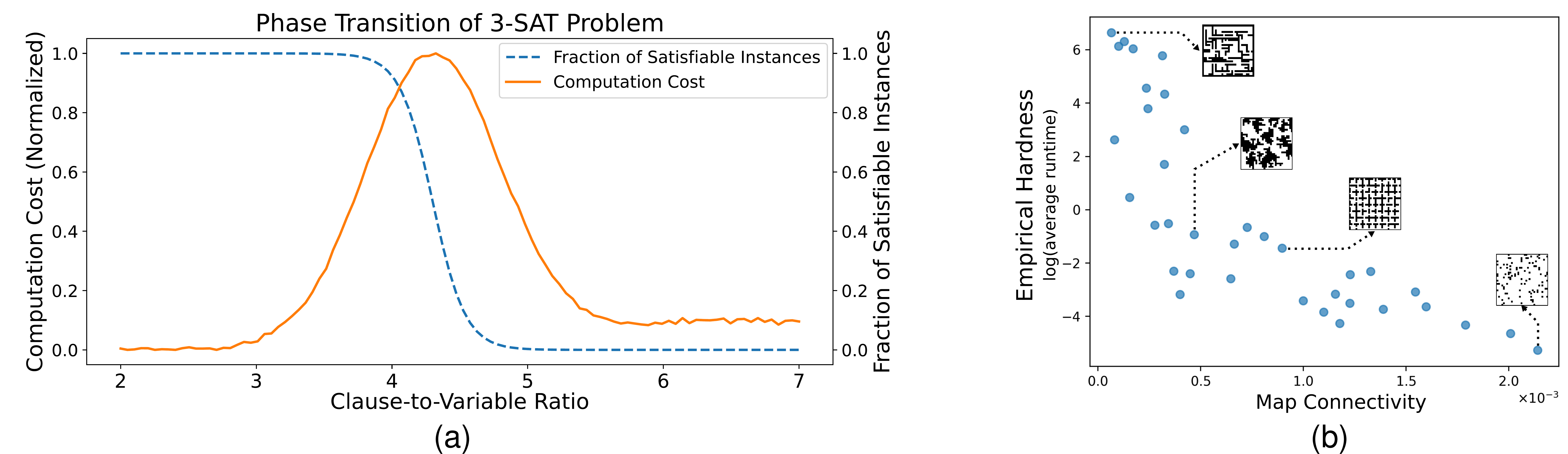}
    \caption{(a) Phase transition of 3-SAT problem. (b) The average empirical hardness of MAPF instances on different maps.}
    \label{fig:phase}
    \Description{Two figures, one showing the phase transition phenomena of the 3-SAT problem. Another figure shows that the average empirical hardness of different layouts of maps can be quite different.}
\end{figure*}

\subsection{Algorithm Selection}

Given a specific instance, algorithm selection aims to choose the best algorithm from a predefined portfolio while minimizing performance objectives such as runtime~\cite{rice1976algorithm}. Algorithm selection has been successfully applied to many computationally hard problems, including SAT~\cite{nudelman2004understanding} and Mixed Integer Programming (MIP)~\cite{xu2011hydra}. Modern approaches use machine learning techniques and treat algorithm selection as a classification or regression problem. 

Algorithm selection has been successfully applied to MAPF problems as well.
There are two major approaches to solving this problem. One is to formulate MAPF algorithm selection as an image classification problem~\cite{sigurdson2019automatic, ren2021mapfast, alkazzi2022mapfaster}. By encoding the MAPF instances as RGB images, a convolutional neural network~(CNN) can be trained to predict the fastest algorithm. The second approach manually selects MAPF instance features, such as obstacle density and agents' distance to goals~\cite{kaduri2020algorithm}, and trains a tree-based model based on these features~\cite{chen2016xgboost}. A recent work~\cite{shabalin2024algorithm} combined hand-crafted features with graph-based encodings and achieved notable improvements.

\textbf{Challenges in Instance Encoding:} The primary challenge in achieving effective algorithm selection lies in encoding MAPF instances accurately. 
Early CNN-based models encoded MAPF instances as RGB images with start and goal locations depicted as pixels in different colors~\cite{sigurdson2019automatic}. However, this approach treated start and goal locations anonymously, failing to distinguish instances with different start-goal permutations and resulting in limited performance. One approach to address this limitation is to add single-agent shortest paths as an additional encoding, which captures both the agent distribution and map topology information~\cite{ren2021mapfast}. This technique has been widely adopted by many CNN-based MAPF algorithm selectors~\cite{alkazzi2022mapfaster, chen2024no, shabalin2024algorithm}. However, there is still room for improvement in encoding strategies, as existing methods may not fully capture the complexity of MAPF instances.

\textbf{Challenges in Feature Selection: }
For feature-based algorithm selectors, it is critical to include the most important instance features that affect the MAPF empirical hardness. But even identifying these instance features is a challenging task. For example, Kaduri et al.~\cite{kaduri2020algorithm} proposed features like agent sparsity and average goal distance. However, these features fail to consider map topology, which also has major impacts on the empirical hardness of MAPF instances~(e.g., Figure~\ref{fig:diff_instances}(b)). Moreover, there is no guarantee that these manually selected instance features cover all important factors.

\subsection{Algorithm Configuration}
Instead of selecting different algorithms, algorithm configuration optimizes the parameter settings for a given algorithm to minimize runtime or maximize solution quality~\cite{hutter2011sequential}. These parameters can be in various forms such as categorical, boolean, or continuous, as long as they affect the algorithm's behavior and performance.

\textbf{Challenges in Heuristic Configurations:} Algorithm configuration remains unexplored in MAPF, as most MAPF solvers such as CBSH2-RTC~\cite{li2021pairwise} or BCP~\cite{lam2019branch} have relatively fewer heuristics to consider compared to other combinatorial problems. However, heuristic configurations still play a crucial role. For example, Figure~\ref{fig:diff_instances}(c) demonstrates that different heuristic combinations for CBSH2-RTC~\cite{li2021pairwise} can lead to significantly varied runtimes. Understanding why certain configurations perform poorly in specific instances could also help develop better heuristics.

\subsection{Future Directions}
Future research should address two key areas:

\textbf{Better Encoding Techniques: } Chen et al.~\cite{chen2024no} found that similar encoding methods only bring limited performance gains in different deep learning-based algorithm selectors. Future research should explore instance encoding techniques beyond single-agent shortest paths to better capture the inherent problem hardness. 

\textbf{Algorithm Configuration: } Optimizing heuristic settings for MAPF solvers across various instances can improve performance. Additionally, analyzing poor configurations could inspire novel heuristics, enhancing solver capabilities and adaptability.

\section{Challenge II: Understanding the Empirical Hardness of MAPF}

Algorithm selection offers practical tools for predicting the best algorithm for each instance but does not explain the factors that cause the variance in instance hardness. For MAPF, understanding these factors is particularly challenging since both map topology and agent distribution will affect empirical hardness.

Empirical hardness is influenced by structural properties that are not immediately apparent from problem size alone. Research in related domains identifies phase transitions, backbone, and backdoors as critical predictors of instance hardness.

\subsection{Phase Transition}\label{sec:phase}

Phase transition describes abrupt changes in system behavior or properties when critical parameters are altered~\cite{sole2011phase}. This phenomenon is widely studied in physics and often describes the transition of different states for substances (e.g., water changes from liquid to gas). Phase transition also exists in many computationally hard problems, where instance hardness shifts sharply at specific thresholds~\cite{cheeseman1991really}. For instance, as shown in Figure~\ref{fig:phase}(a), the computation cost for 3-SAT instance exhibits an \textit{easy-hard-easy}~\cite{cheeseman1991really} pattern as the clause-to-variable ratio changes. The computation cost peaked as the fraction of satisfiable instances shifted from satisfiable to unsatisfiable. This discovery has enabled the systematic generation of challenging SAT instances by fine-tuning the clause-to-variable ratio ~\cite{selman1996generating}. Similar transitions have been observed in other NP-complete problems, such as graph coloring~\cite{achlioptas1999sharp}, Hamiltonian circuits~\cite{cheeseman1991really}, and traveling salesman~(TSP)~\cite{gent1996tsp}.

Phase transitions are particularly valuable for generating benchmark instances to evaluate algorithm performance. By tuning parameters near critical thresholds, researchers can systematically create hard instances~\cite{selman1996generating}. Exploring whether phase transitions also exist in the MAPF problem is undoubtedly a fascinating research topic. However, studying phase transitions in non-decision problems, like optimization tasks, is more challenging since they lack clear solvability thresholds. 

\textbf{Challenges in MAPF Phase Transition:}
Studying the phase transition for MAPF problems is particularly challenging, as MAPF is more complex than SAT problems. It is very hard to find a single numerical parameter as clause-to-variable ratio in SAT that correlates with problem hardness. Using numerical parameters to capture map topology information is also very challenging. 

\textbf{Challenges in Map Topology and Agent Distribution:} As shown in Figure~\ref{fig:diff_instances}(a)(b), both map topology and agent distribution can affect the empirical hardness. To address this, one approach is to assume agents are uniformly distributed and focus on map features and average empirical hardness. Ewing et al.~\cite{ewing2022betweenness} showed maps with higher betweenness centrality tend to have harder instances. Moreover, as shown in Figure~\ref{fig:phase}(b), poorly connected maps generally exhibit higher average empirical hardness compared to well-connected ones~\cite{ren2024map}. While these studies provide a foundation for exploring MAPF phase transitions, removing the assumptions on agent distribution remains challenging.

\subsection{Backbone and Backdoor}
Backbone and backdoor are important concepts to understand empirical hardness, offering insights into problem instance structures. 

\textbf{Backbone: } The backbone represents a set of variables whose values are identical across all possible solutions of an instance~\cite{monasson1999determining}. The backbone is often used as a key indicator for empirical hardness. In SAT problems, a large number of backbone variables often indicate an over-constrained instance and it is hard to find feasible solutions~\cite{achlioptas2000generating, parkes1997clustering}. Early identification and elimination of the backbones will help improve the algorithm's performance~\cite{monasson1999determining, dubois2001backbone}.

\textbf{Backdoor: } The backdoor represents a set of variables that, when properly assigned, simplifies the problem~\cite{williams2003backdoors}. The size of the backdoor is an effective indicator for estimating empirical hardness, as smaller backdoors indicate well-structured and easier instances~\cite{kilby2005backbones, ruan2004backdoor}. For example, setting the backdoor variables properly for SAT instances will make the remaining formula solvable in polynomial time by certain SAT solvers~\cite{williams2003connections}. 

\textbf{Challenges in Backbone/Backdoor:}
There is no existing research on the backbone/backdoor of MAPF problems. With MAPF's complex instance features compared to SAT, the main challenge is how to properly define the backbone/backdoor. In a broader concept, the backbone could be caused by specific structures that dominate the map connectivity, such as narrow corridors or choker points. These structures might force agents to traverse certain locations in all possible solutions, causing more conflicts between agents, and making the instances harder. The backdoor could be a subset of agents causing the majority of conflicts. Planning their paths first may simplify the remaining problem.

\subsection{Future Directions}
Future research should focus on investigating phase transitions and backbone/backdoor, such as exploring different types of phase transitions for MAPF or providing formal definitions of backbone and backdoor. These studies can provide systematic insights into instance hardness and guide the development of novel algorithms.

\section{Challenge III: Generating Hard MAPF Instances and Benchmark Datasets}
Once the key instance features influencing MAPF empirical hardness are identified, the next challenge is how to effectively leverage this knowledge to generate hard instances. With the development of increasingly powerful algorithms~\cite{gange2019lazy, li2021pairwise, lam2022branch}, there is also a growing need for more challenging and diverse benchmark datasets to thoroughly evaluate their performance. 

\subsection{Generating Hard MAPF Instances}
Generating hard instances is not an easy task, especially without a profound understanding of empirical hardness. For example, randomly generated SAT instances are generally easy to solve, but hard instances occur more frequently at specific clause-to-variable ratios~\cite{selman1996generating}.
The study of phase transition has inspired numerous methods of generating hard instances, as the most challenging instances often exist in phase transition regions. \citeauthor{selman1996generating}~\cite{selman1996generating}
demonstrated that challenging SAT instances can be generated by setting the clause-to-variable ratio to approximately $4.25$.

\textbf{Challenges in Hard Instance Generation: }Since phase transition in MAPF has not been formally studied, it cannot be used to guide the generation of hard instances as it does in other research domains. Nevertheless, MAPF researchers have explored various alternative methods for creating challenging instances. The key challenge is that both agent distribution and map topology will affect the empirical hardness of MAPF instances. And it is very hard to study them together.

Some researchers make assumptions about the distribution of agents' start and goal locations and focus on the topology of maps. The most common assumption is that agents' start and goal locations are sampled uniformly at random. This scenario studies the average empirical hardness of the MAPF problem on different maps. For instance, \citeauthor{ewing2022betweenness}~\cite{ewing2022betweenness} showed that the empirical hardness is correlated with the betweenness centrality of the maps. Betweenness centrality measures how often a map cell is traversed by the shortest paths and a high betweenness centrality often indicates choker points. Similarly, Ren et al.~\cite{ren2024map} showed that hard instances are more common on poorly connected maps and demonstrated using the Quality Diversity~(QD) method to generate maps with controllable connectivity. However, as discussed in Section~\ref{sec:phase}, removing the assumptions on agent distribution remains challenging.

\subsection{Generating Better Benchmark Datasets}

Developing better benchmark datasets with a wide range of hardness is crucial for evaluating MAPF algorithms. Research that initially focused on simplifying instances has provided valuable insights. For instance, \citeauthor{zhang2024multi}~\cite{zhang2024multi} used QD methods to optimize warehouse throughput, which can be viewed as indirectly generating maps with lower average empirical hardness. This approach was improved with NCA~\cite{mordvintsev2020growing} and CMA-MAE~\cite{fontaine2023covariance} for scalability~\cite{zhang2024arbitrarily}. \citeauthor{qian2024quality}~\cite{qian2024quality} later proposed a more powerful framework for generating diverse MAPF datasets with varying hardness.

\textbf{Challenges in Benchmarking: }The key challenge is to choose the best metrics for quality diversity methods to generate diverse instances. \citeauthor{qian2024quality}~\cite{qian2024quality} identified obstacle density and the Kullback–Leibler (KL) divergence of tile patterns as effective indicators for general hardness and spatial arrangement. While these metrics provide a foundation for benchmark generation and rigorous MAPF solver evaluation, many other potential metrics remain unexplored.

\subsection{Future Directions}
Generating hard MAPF instances without general agent distribution assumptions remains challenging. Possible directions are:  

\textbf{Reinforcement Learning:} Train an instance generator with a reward function that encourages conflicts between agents, making instances harder for conflict-based algorithms.

\textbf{Specialized Hard Instances:} Extract MAPF instances from other problems. For example, one could extract challenging SAT instances from a MAPF problem~\cite{surynek2019unifying}. The main challenge is identifying complex problems that can be effectively reduced to MAPF.


\section{Conclusion}
This paper identifies and presents three key research challenges in understanding and leveraging the empirical hardness of multi-agent pathfinding (MAPF). First, selecting the best-performing algorithm for a given instance remains an open problem, requiring advancements in better instance encoding techniques in algorithm selection and configuration. Second, understanding the factors that influence instance hardness, such as phase transitions and other structural features, is essential for building a deeper theoretical foundation. Finally, leveraging this knowledge to generate challenging MAPF instances and diverse benchmark datasets can lead to the development of more robust and efficient algorithms. Addressing these challenges will bridge the gap between theoretical complexity and practical hardness and ultimately advance our understanding of the true essence of the MAPF problem.



\begin{acks}
This research was supported by the National Science Foundation (NSF) under grant 2330942 at Brown University, 2112533 at the University of Southern California, and 2434916, 2346058, 2321786, 2121028, and 1935712 at the University of California, Irvine, as well as gifts from Amazon Robotics. The views and conclusions contained in this document are those of the authors and should not be interpreted as representing the official policies, either expressed or implied, of the sponsoring organizations, agencies, or the U.S. government.

\end{acks}



\bibliographystyle{ACM-Reference-Format} 
\bibliography{mybib}


\end{document}